\documentstyle[]{mn}

\def\kms{ $\rmn{km~s^{-1}}$}

\newif\ifAMStwofonts 
\ifoldfss 
  \newcommand{\rmn}[1] {{\rm #1}}

  \ifCUPmtlplainloaded \else 
    \NewTextAlphabet{textbfit} {cmbxti10} {} 
    \NewTextAlphabet{textbfss} {cmssbx10} {} 
    \NewMathAlphabet{mathbfit} {cmbxti10} {} 
    \NewMathAlphabet{mathbfss} {cmssbx10} {} 
  \fi 
  \ifAMStwofonts 
    \ifCUPmtlplainloaded \else 
      \NewSymbolFont{upmath} {eurm10} 
      \NewSymbolFont{AMSa} {msam10} 
      \NewMathSymbol{\upi}     {0}{upmath}{19} 
      \NewMathSymbol{\umu}     {0}{upmath}{16} 
      \NewMathSymbol{\upartial}{0}{upmath}{40} 
      \NewMathSymbol{\leqslant}{3}{AMSa}{36} 
      \NewMathSymbol{\geqslant}{3}{AMSa}{3E}

      \let\leq=\leqslant  
        
    \fi 
  \fi 
\fi 
\ifnfssone 
  \newmathalphabet{\mathit} 
  \addtoversion{normal}{\mathit}{cmr}{m}{it} 
  \addtoversion{bold}{\mathit}{cmr}{bx}{it} 
  \newcommand{\rmn}[1] {\mathrm{#1}}

  \newmathalphabet{\mathbfit} 
  \addtoversion{normal}{\mathbfit}{cmr}{bx}{it} 
  \addtoversion{bold}{\mathbfit}{cmr}{bx}{it} 
  \newmathalphabet{\mathbfss} 
  \addtoversion{normal}{\mathbfss}{cmss}{bx}{n} 
  \addtoversion{bold}{\mathbfss}{cmss}{bx}{n} 
  \ifAMStwofonts 
    \ifCUPmtlplainloaded \else 
      \UseAMStwoboldmath 
      \makeatletter 
      \new@mathgroup\upmath@group 
      \define@mathgroup\mv@normal\upmath@group{eur}{m}{n} 
      \define@mathgroup\mv@bold\upmath@group{eur}{b}{n} 
      \edef\UPM{\hexnumber\upmath@group} 
      \new@mathgroup\amsa@group 
      \define@mathgroup\mv@normal\amsa@group{msa}{m}{n} 
      \define@mathgroup\mv@bold\amsa@group{msa}{m}{n} 
      \edef\AMSa{\hexnumber\amsa@group} 
      \makeatother 
      \mathchardef\upi="0\UPM19 
      \mathchardef\umu="0\UPM16 
      \mathchardef\upartial="0\UPM40 
      \mathchardef\leqslant="3\AMSa36 
      \mathchardef\geqslant="3\AMSa3E 

      \let\leq=\leqslant  

    \fi 
  \fi 
\fi 

\ifnfsstwo 
  \newcommand{\rmn}[1] {\mathrm{#1}}

  \DeclareMathAlphabet{\mathbfit}{OT1}{cmr}{bx}{it} 
  \SetMathAlphabet\mathbfit{bold}{OT1}{cmr}{bx}{it} 
  \DeclareMathAlphabet{\mathbfss}{OT1}{cmss}{bx}{n} 
  \SetMathAlphabet\mathbfss{bold}{OT1}{cmss}{bx}{n} 
  \ifAMStwofonts 
    \ifCUPmtlplainloaded \else 
      \DeclareSymbolFont{UPM}{U}{eur}{m}{n} 
      \SetSymbolFont{UPM}{bold}{U}{eur}{b}{n} 
      \DeclareSymbolFont{AMSa}{U}{msa}{m}{n} 
      \DeclareMathSymbol{\upi}{0}{UPM}{"19} 
      \DeclareMathSymbol{\umu}{0}{UPM}{"16} 
      \DeclareMathSymbol{\upartial}{0}{UPM}{"40} 
      \DeclareMathSymbol{\leqslant}{3}{AMSa}{"36} 
      \DeclareMathSymbol{\geqslant}{3}{AMSa}{"3E} 

      \let\leq=\leqslant  

    \fi 
  \fi 
\fi 

\ifCUPmtlplainloaded \else 
  \ifAMStwofonts \else 
    \def\upi{\pi} 
    \def\umu{\mu} 
    \def\upartial{\partial} 
  \fi 
\fi 
 
\input epsf.sty 

\title[Multiple stellar systems -- III. HD 109648] 
{Studies of multiple stellar systems -- III. Modulation of orbital elements 
in the triple-lined system HD 109648}
\author[S. Jha et al.]
{Saurabh Jha,$^1$ Guillermo Torres,$^1$ Robert P. Stefanik,$^1$ David W.
Latham$^1$
\newauthor and Tsevi Mazeh$^2$
\thanks{E-mail:~\{sjha, gtorres, rstefanik, dlatham\}@cfa.harvard.edu;
mazeh@wise.tau.ac.il} \\
$^1$Harvard-Smithsonian Center for Astrophysics, 60 Garden Street, Cambridge,
MA 02138, USA\\
$^2$School of Physics and Astronomy, Raymond and Beverly Sackler Faculty of
Exact Sciences, Tel Aviv University, Tel Aviv, Israel}
\date{submitted September 22, 1999; accepted March 30, 2000}

\pagerange{\pageref{firstpage}--\pageref{lastpage}}
\pubyear{2000}

\begin{document}

\maketitle

\label{firstpage}

\begin{abstract}
The triple-lined spectroscopic triple system HD 109648 has one of the
shortest periods known for the outer orbit in a late-type triple,
120.5 days, and the ratio between the periods of the outer and inner
orbits is small, 22:1.  With such extreme values, this system should
show orbital element variations over a timescale of about a decade.
We have monitored the radial velocities of HD 109648 with the CfA
Digital Speedometers for eight years, and have found evidence for
modulation of some orbital elements.  While we see no definite evidence for
modulation of the inner binary eccentricity, we clearly observe variations in
the inner and outer longitudes of periastron, as well as in the radial
velocity amplitudes of the three components. The observational
results, combined with numerical simulations, allow us to put
constraints on the orientation of the orbits.
\end{abstract}


\section{Introduction}

The number of triple systems with well-determined orbital elements is
still small (Fekel 1981; Tokovinin 1997, 1999).  In particular, the
number of spectroscopic triples in which the wide orbit is also known
from radial-velocity observations is very small.  Part of the problem
is that the velocity amplitude of the outer binary is usually small
compared to the amplitude of the inner binary.  Moreover, after a
binary orbit has been solved, the natural reaction is to discontinue
observing it, and checks for longer-term variations are rarely made
\cite{may87}.  This series of papers is aimed at increasing our
knowledge of triples by investigating systems where the inner and
outer orbits can both be determined from spectroscopic observations.
The first paper of the series (Mazeh, Krymolowski and Latham 1993,
hereinafter Paper I) examined the halo triple G38-13, while the second
paper (Krymolowski and Mazeh 1998, hereinafter Paper II) derived an
analytic technique which allows for fast simulation of orbital
modulations of a binary induced by a third star.

In the present paper we analyse the triple-lined spectroscopic triple
system HD 109648 (HIP 61497, $\alpha = 12^{\rmn{h}}35^{\rmn{m}}59 \fs
8$, $\delta = +36\degr 15\arcmin 30\arcsec$ (J2000); $V = 8.8$).  HD
109648 was identified as one (star 6) of a handful of stars belonging
to the remnant of a nearby old open cluster, Upgren~1 \cite{upg65},
but subsequent studies have weakened the interpretation that all of
the stars originally identified are physically associated (Upgren,
Philip \& Beavers 1982; Gatewood et al.\ 1992; Stefanik et al.\ 1997;
Baumgardt 1998).

The triple-lined nature of HD 109648 was noticed soon after we began
observing it, because the one-dimensional correlations of some of the
spectra clearly showed three peaks. A periodicity analysis revealed
periods at $\sim 5.5$ and $\sim 120$ days. Triple systems tend to be
hierarchical, usually with a close binary and a more distant third
star, as other configurations are generally unstable and are unlikely
to persist and be detected. To first-order, a hierarchical triple
system can be separated into an inner orbit (comprising the two close
stars) and an outer orbit (comprising the third star and the
centre-of-mass of the inner pair). This approximation is most valid
when the distance to the third star far exceeds the separation between
the inner two stars. One of the goals of this study is to investigate
the interaction of these three stars (through the variation of the
inner and outer orbits) to higher order.

A preliminary version of this work was presented at a conference
entitled `Thirty Years of Astronomy at the Van Vleck Observatory: A
Meeting in Honor of Arthur R. Upgren' \cite{jha97}. This paper updates
the orbital solutions presented there and adds a significantly more
detailed analysis of the system, partly through the use of numerical
simulations.

In Section 2 we summarise the analysis of the observations, including
the derivation of the radial velocities, orbital solutions, and
additional parameters such as the mass ratios and constraints on the
orbital inclinations.  We discuss the theoretically expected
modulations of orbital elements in Section 3. In Section 4 we describe
our efforts to search for such variations and present our
results. Further constraints for the system via numerical simulation
are derived in Section 5. Finally, in Section 6 we discuss our results
and relate them to previous and future work.

\section{Radial velocities and orbital solutions}
 
HD 109648 has been monitored since 1990 with the Center for
Astrophysics (CfA) Digital Speedometer \cite{lat85,lat92} on the 1.5-m
Wyeth Reflector at the Oak Ridge Observatory, located in the town of
Harvard, Massachusetts. The echelle spectra cover 45 \AA\ centered at
5187 \AA, with a spectral resolution of $\lambda/\Delta \lambda
\approx 35,000$. As of 1998, we have secured 290 spectra of HD109648.

Radial velocities were derived for each of the three stars in
the system using the three-dimensional version \cite{zuc95} of the
two-dimensional correlation technique TODCOR \cite{zuc94}.  TODCOR
assumes that the spectrum for each individual star in the system is
known, and that an observed spectrum is composed of the individual
component spectra added together, each shifted by its own radial
velocity.  Thus, to use TODCOR successfully one must have suitable
template spectra for each of the components.  We chose our templates
from a grid of synthetic spectra calculated by Jon Morse using the 1992
Kurucz model atmospheres (e.g. Nordstr\"{o}m et al. 1994).

Our first guess for the template parameters was based on a visual inspection
of the spectra. Application of TODCOR yielded preliminary velocities for each
of the three components, from which we determined the makeup of the triple:
the inner binary consists of the primary and the tertiary, while the outer
star is second in brightness. With this information, we were able to refine
our templates to obtain the final velocities. For the primary we adopted an
effective temperature, $T_{\rm eff} = 6750$ K; solar metallicity, [m/H] = 0;
and main-sequence surface gravity, log $g = 4.5$ (cgs units). The period of
the inner binary is quite short, and we assumed this has led to spin-orbit
synchronization for the inner stars (see for example Mazeh \& Shaham
1979). Therefore we adopted a rotational velocity of $v \sin i = 10$ \kms\ for
both of the inner binary stars. For the secondary and the tertiary, we have
used a slightly cooler temperature, $T_{\rm eff} = 6500$ K, and for the outer
star we assumed that the rotation was negligible, $v \sin i = 0$\kms. Small
changes in the choice of template parameters did not have much effect on the
radial velocities or orbital solutions. The final template parameters are
listed in Table \ref{tab-temparm}, and the individual radial velocities are
reported in the Appendix.

\begin{table}
\caption{Parameters for the synthetic template spectra.}
\begin{tabular}{ccccc}
Star & $T_{\rmn{eff}}$ & $\log g$ & [m/H] & $v_{\rmn{rot}}$ \\
     &       K         & $\log(\rmn{cm~s^{-2}})$ & dex & $\rmn{km~s^{-1}}$ \\
\hline
primary (Aa) & 6750 & 4.5 & 0.0 & 10 \\
secondary (B)  & 6500 & 4.5 & 0.0 &  0 \\
tertiary (Ab) & 6500 & 4.5 & 0.0 & 10 \\
\hline
\end{tabular}
\label{tab-temparm}
\end{table}

\begin{table} 
\caption{HD 109648 mean orbital solution.} 
\begin{tabular}{lll} 
\hline 
$P_A$                   & $5.4784499 \pm 0.0000080$     & days \\
$K_{Aa}$                & $62.02 \pm 0.11$              & $\rmn{km~s^{-1}}$ \\
$K_{Ab}$                & $68.98 \pm 0.16$              & $\rmn{km~s^{-1}}$ \\
$e_A$                   & $0.0119 \pm 0.0014$           &  \\
$\omega_A$              & $37.2 \pm 7.1$                & deg \\
$T_A$                   & $2448462.24 \pm 0.11$         & HJD \\
\\
$a_{Aa}~\sin i_A$       & $4.6721 \pm 0.0084$           & Gm \\
$a_{Ab}~\sin i_A$       & $5.1965 \pm 0.0122$           & Gm \\
$m_{Aa}~\sin^3 i_A$     & $0.6719 \pm 0.0034$           & $\rmn{M_{\sun}}$ \\
$m_{Ab}~\sin^3 i_A$     & $0.6041 \pm 0.0026$           & $\rmn{M_{\sun}}$ \\
\\
$P_{AB}$                & $120.5275 \pm 0.0080$         & days\\
$K_A$                   & $17.20 \pm 0.10$              & $\rmn{km~s^{-1}}$ \\
$K_B$                   & $34.91 \pm 0.15$              & $\rmn{km~s^{-1}}$ \\
$e_{AB}$                & $0.2362 \pm 0.0033$           &  \\
$\omega_{AB}$           & $331.79 \pm 0.84$             & deg \\
$T_{AB}$                & $2448412.19 \pm 0.28$         & HJD \\
\\
$a_A~\sin i_{AB}$       & $27.70 \pm 0.15$              & Gm \\
$a_B~\sin i_{AB}$       & $56.23 \pm 0.24$              & Gm \\
$m_A~\sin^3 i_{AB}$     & $1.0862 \pm 0.0114$           & $\rmn{M_{\sun}}$ \\
$m_B~\sin^3 i_{AB}$     & $0.5352 \pm 0.0058$           & $\rmn{M_{\sun}}$ \\
\\
$\gamma$                & $-18.900 \pm 0.055$           & $\rmn{km~s^{-1}}$ \\
\\
$\sigma(O-C)_{Aa}$      & $1.36$                        & $\rmn{km~s^{-1}}$\\
$\sigma(O-C)_{Ab}$      & $1.97$                        & $\rmn{km~s^{-1}}$\\
$\sigma(O-C)_{B}$       & $1.72$                        & $\rmn{km~s^{-1}}$\\
\hline
\end{tabular}
\label{tab-elem}
\end{table}

For a proper solution of the orbital elements, one must solve for the inner
and outer motions simultaneously. For this purpose we have used {\sc orb20}, a
code developed at Tel Aviv University \cite{pap1}, and a new code developed at
the CfA.  These two independent codes yielded the same results. Throughout
this paper we employ the following notation for the elements of this
hierarchical triple system.  We label the inner stars Aa and Ab (with Aa being
the brighter), while the centre-of-mass of the inner stars is denoted as A and
the outer star is denoted B. When we are discussing orbits rather than the
stars themselves, we designate the inner orbit as A and the outer orbit as AB.

\begin{figure}
\def\epsfsize#1#2{0.46#1}
\epsfbox{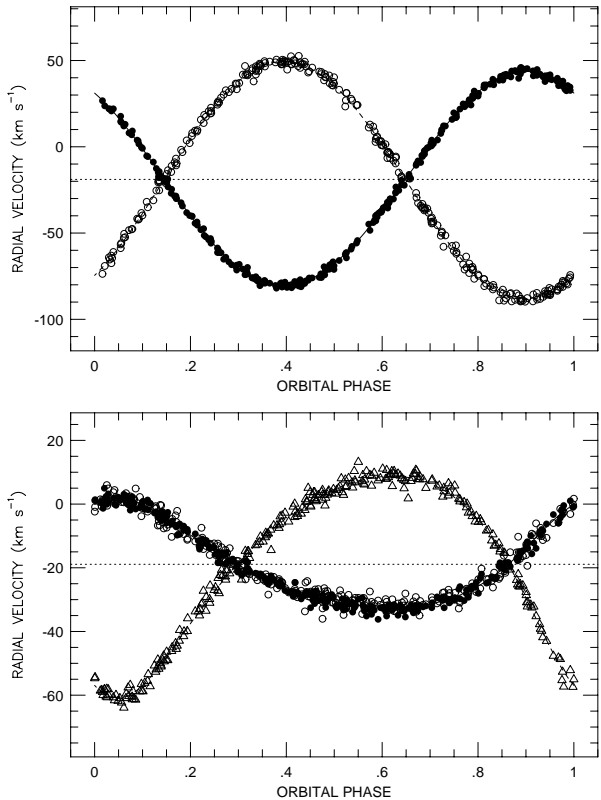}
\caption{Simultaneous orbital solution for HD 109648 using all of the 290
observations. The upper panel shows the stars of the inner binary, Aa
(filled circles) and Ab (empty circles), with their centre-of-mass
motion removed. The lower panel shows the outer binary, comprised of
the outer star B (triangles), as well as the common centre-of-mass
motion of the inner stars. The observations have
velocity residuals $\sigma_{Aa} = 1.36$, $\sigma_{Ab} = 1.97$
and~$\sigma_{B} = 1.72 \; \rmn{km~s^{-1}}$, part of which
arises from the fact the orbital elements are not static.}
\label{fig-orb}
\end{figure}

The orbital solution using all 290 of our observations is
displayed in Figure \ref{fig-orb}. The top panel shows the motion of
the two inner stars, after their centre-of-mass motion has
been removed. The bottom panel shows the centre-of-mass motion of the
inner binary, as well as the motion of the third star. The derived
average orbital elements of the inner and outer motions, as well as
the overall radial velocity of the system, $\gamma$, are listed in
Table \ref{tab-elem}.

The triple-lined nature of the spectra yields much more information about the
system than can be gathered when only one component is visible (as was the case
for G38-13 in Paper I). We can, for instance, determine the mass ratios between
the three stars. The mass ratio of the inner pair is easily determined
from the inner orbital elements, $m_{Aa} K_{Aa} = m_{Ab} K_{Ab}$. This
results in
\begin{equation}
\frac{m_{Ab}}{m_{Aa}} = 0.8991 \pm 0.0027.
\end{equation}
The same relation holds for the outer orbit, $m_{A} K_{A} = m_{B}
K_{B}$, but in this triple system we know that $m_{A} = m_{Aa} +
m_{Ab}$. Thus we derive
\begin{equation}
\frac{m_{B}}{m_{Aa}} = \frac{K_A}{K_B}\left(1 +
\frac{K_{Aa}}{K_{Ab}}\right) = 0.9356 \pm 0.0072.
\end{equation}
From the orbital elements and Kepler's Third Law we can also derive
the quantities $m_{Aa} \sin^3 i_A$ and $m_{B} \sin^3 i_{AB}$ (e.g., Batten
1973) listed in Table \ref{tab-elem}, which with the mass ratio lead
to a ratio involving the inclination angles,
\begin{equation}
\frac{\sin i_{AB}}{\sin i_A} = 0.9478 \pm 0.0045.
\end{equation}

The individual inclination angles remain unknown, and consequently, so
do the exact masses of the stars, but we can estimate the masses by
assuming the stars are still on the main sequence. Upgren \& Rubin
(1965) report an MK spectral type of F6V for HD 109648. The light of
the primary dominates the spectrum of the system, so this spectral
type corresponds to a primary mass \cite{gra92} of
\begin{equation}
m_{Aa} = 1.3 \pm 0.1 \; M_{\sun},
\end{equation}
where we have assumed solar metallicity and have adopted an uncertainty
that corresponds to the spectral-type range F3V to F8V.
Using the known mass ratios, we can then calculate the masses of the
other two stars.

From the masses we can nearly determine the inclination angles. Since
radial velocity measurements do not reveal the direction of orbital
motion, we are left with an ambiguity between the following
supplementary possibilities for each angle,
\begin{equation}
i_A = 53.4 \pm 2.0 \degr \quad \hbox{or} \quad 126.6 \pm 2.0 \degr 
\qquad \hbox{and}
\end{equation}
\begin{equation}
i_{AB} = 49.5 \pm 1.8 \degr \quad \hbox{or} \quad 130.5 \pm 1.8 \degr.
\end{equation}

However, even if we were to resolve the ambiguities in the
inclinations, we would still not know the complete geometric
orientation of the system. Spectroscopic observations (as opposed to
visual ones) do not provide the elements
$\Omega_{A}$~and~$\Omega_{AB}$, the position angles of the inner and
outer lines of nodes. Without these angles, we cannot determine an
important quantity in the interaction of the two binaries---the
relative inclination $\phi$. The relative inclination is defined as
the angle between the two orbital planes, or identically, the angle
between the inner and outer angular momentum vectors.  It is related
to the individual inclination angles, and the angles of the lines of
nodes, by (Batten 1973; Fekel 1981)
\begin{equation}
\cos \phi = \cos i_A \cos i_{AB} + \sin i_A \sin i_{AB} \cos
(\Omega_A - \Omega_{AB}).
\end{equation}
Since the quantity $\Omega_A - \Omega_{AB}$ is unknown, we can only limit its
cosine between $+1$ and $-1$, resulting in a geometrical constraint on $\phi$,
\begin{equation}
i_A - i_{AB} \leq \phi \leq i_A + i_{AB}.
\end{equation}
Nevertheless, from this result we can derive an important parameter,
namely the minimum relative inclination angle, to see if the system
can be coplanar. We determine that the minimum angle between the
orbital planes is $\phi_{\rmn{min}} = 3.9 \pm 0.3 \degr$. Thus the two
orbits could be very close to coplanarity, but 
cannot be exactly coplanar. In Section 4 we strengthen this
lower limit slightly.

\section{Expected effects of the Three-body interaction} 

As discussed in Papers I and II, the separation of the motions of a
hierarchical triple system into inner and outer orbits is only a
first-order approximation. The gravitational attraction of the outer
body exerted on each of the two inner bodies is different from the
gravitational attraction exerted on an imaginary body at the
centre-of-mass of the inner binary system. The difference induces
long-term modulations of some of the orbital elements of the system.

The timescale for such modulations is on the order of (Mazeh \& Shaham
1979; Paper II)
\begin{equation} 
T_{\rmn{mod}} = P_{AB} \left( \frac{P_{AB}}{P_A} \right) \left(
\frac{m_{Aa}+m_{Ab}}{m_B} \right).
\label{eqn-modtim}
\end{equation} 
Thus, for such modulations to be observationally detectable in a
relatively short time, one requires a system with a short outer
period, as well as a small outer:inner period ratio. HD 109648
satisfies both these requirements, with an outer period about 120.5
days and a period ratio near 22:1. This results in $T_{\rmn{mod}} \sim
15$ years, one of the shortest modulation timescales known for a
late-type triple system. Our observations of HD 109648 span more than
eight years, giving us some hope that we may be able to detect changes
in some of the orbital parameters.

\subsection{Modulation of the inner eccentricity and the longitudes 
of periastron} 

One effect expected from the three-body interaction is a modulation of
the inner binary eccentricity \cite{maz79}. The presence of the third
star causes a quasi-periodic variation in the inner eccentricity,
$e_A$, around an average value. The amplitude of the eccentricity
modulation strongly depends on the eccentricity of the outer orbit
\cite{maz79} and on the relative inclinations between the orbital
planes (Mazeh, Krymolowski \& Rosenfeld 1997; Paper II), with coplanar
situations producing the least effect.

The inner eccentricity modulation goes together with the motions of
the lines of apsides of the two orbits (Mazeh, Krymolowski \&
Rosenfeld 1997; Paper II; Holman, Touma \& Tremaine 1997), which
manifest themselves through the variation of the longitudes of
periastron. Both modulations, that of the inner binary eccentricity
and that of the longitudes of periastron can be observed as changes in
the elements derived for the two orbital motions. 

Mazeh and Shaham (1979) have shown that the inner eccentricity
modulation takes place even when the binary orbit starts as circular
one. This aspect of the eccentricity modulation is applicable here,
because we expect a binary with a period of about 5.5 days to be
completely circularized (Zahn 1975; Mathieu and Mazeh 1984), if it
were not for the effect of the third star.

\subsection{Precession of the nodes} 

Another expected modulation results from an effect known as the
precession of the nodes \cite{maz76}. In the general case of a
non-coplanar triple, the inner and outer angular momentum vectors
($\bmath{G_A}$\ and\ $\bmath{G_{AB}}$) precess around their sum, the
total angular momentum ($\bmath{G}$), which remains fixed. As a result,
the angle between $\bmath{G_A}$ (or $\bmath{G_{AB}}$) and any fixed
direction in space (other than that coincident with $\bmath{G}$), varies
periodically with time.

The observer's line of sight is one such fixed direction. The angle
between the line of sight and $\bmath{G_A}$ is precisely the inner
inclination angle $i_A$, since $\bmath{G_A}$ is perpendicular to the
(instantaneous) inner orbital plane. Similarly, the angle between the
line of sight and $\bmath{G_{AB}}$ is the outer inclination angle
$i_{AB}$. As a result of the precession of the orbital planes we
expect to see a periodic modulation of the inner and outer inclination
angles. In the case of a fixed relative inclination between the two
orbits, the time variations of the inclination angles are given by
\cite{maz76}
\begin{equation}
\cos i_A = \cos \alpha \cos \beta_A + \sin \alpha \sin \beta_A \cos
[\omega_p (t - t_0)] \qquad \hbox{and}
\label{eqn-cosia}
\end{equation}
\begin{equation}
\cos i_{AB} = \cos \alpha \cos \beta_{AB} - \sin \alpha \sin \beta_{AB} \cos
[\omega_p (t - t_0)],
\label{eqn-cosiab}
\end{equation}
where $\alpha$ is the angle between the line of sight and $\bmath{G}$,
$\beta_A$ is the angle between $\bmath{G_A}$ and $\bmath{G}$,
$\beta_{AB}$ is the angle between $\bmath{G_{AB}}$ and $\bmath{G}$,
$\omega_p$ is the angular precession frequency and $t_0$ is a fiducial
time determining the phase. An approximate expression for the
precession frequency $\omega_p$ is given by Mazeh \& Shaham
\shortcite{maz76}; in general it corresponds to the typical modulation
timescale given in equation \ref{eqn-modtim}. The amplitude of the
modulations in the inclination angles are set by $\alpha$, $\beta_A$
and $\beta_{AB}$, and the variations of $i_A$ and $i_{AB}$ are exactly
out of phase.

The modulation of the inclination angles has an immediately observable
effect, because the observed amplitudes of the radial velocity
variations $K$ in a binary system are directly proportional to $\sin
i$ (Mazeh \& Shaham 1976; Mazeh and Mayor 1983). Thus, for HD 109648,
periodic modulations in the inner inclination angle, $i_A$, lead to
periodic modulations in $K_{Aa}$\ and\ $K_{Ab}$. Correspondingly, the
modulations of $i_{AB}$ would be evidenced by variations in $K_A$ and
$K_B$.

\section{Search for Modulations Induced by the Third Star}

\begin{figure}
\def\epsfsize#1#2{0.95#1}  
\epsfbox{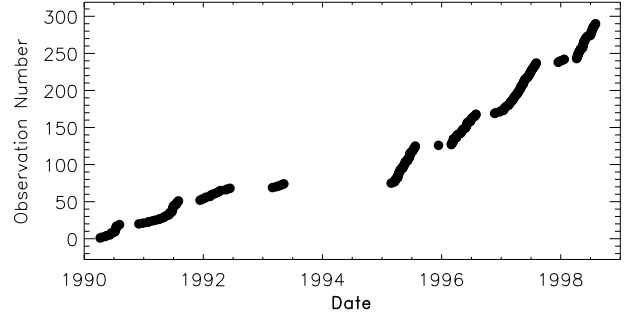} 
\caption{Time history of the 290 observations spanning
more than 3000 days.}
\label{fig-timhist} 
\end{figure}

To search for evidence of modulation of the orbital elements, we have divided
our data set and performed orbital solutions on each subset. To obtain a
robust orbital solution, we would like as many points in each subset as
possible; however, to resolve changes in the elements with time, we would like
as many subsets as possible. The ultimate constraint comes from the time
history of our observations, shown in Figure \ref{fig-timhist}, which has
forced us to use only five subsets. Data from the first few years of
observation (before we appreciated the importance of getting good coverage of
this system) were combined into one subset, while the observations from each
subsequent season form their own subset. We have tried further divisions of
the data (e.g., separating the first subset into two), but these provided
orbital solutions which were too uncertain to be useful. As expected (Paper
II), the inner and outer periods did not vary over the timespan of the
observations. Because of this and the fact that our later subsets cover less
than two outer periods, in what follows we have fixed the outer period at the
value determined by using all the observations.

\subsection{Modulation of the inner eccentricity and the longitudes of
periastron} 

\begin{figure}
\def\epsfsize#1#2{0.95#1}
\epsfbox{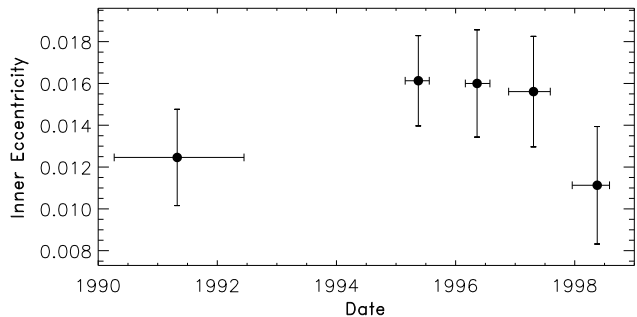}
\caption{The inner eccentricity as a function of time. }
\label{fig-evar}
\end{figure}

\begin{figure*} 
\def\epsfsize#1#2{0.95#1}  
\epsfbox{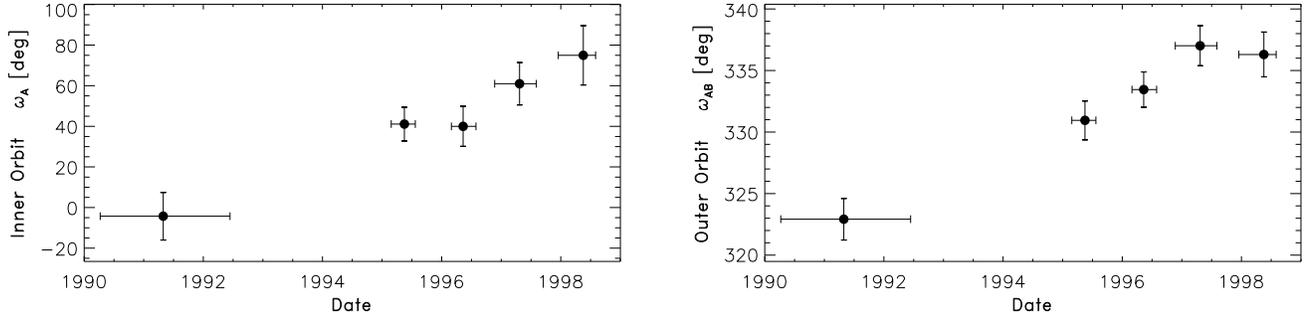} 
\caption{Variation of the inner and outer longitudes of periastron,
$\omega_{A}$ and $\omega_{AB}$.} 
\label{fig-omvar} 
\end{figure*}

The inner eccentricity and the longitudes of periastron are presented
as a function of time in Figure \ref{fig-evar} and Figure
\ref{fig-omvar}.  In these and subsequent figures, the horizontal
``error bars'' actually indicate the time span of each subset, while
the plotted points are at the mean date within the subset (each
observation was given equal weight). There is no obvious modulation of
the inner eccentricity, and this fact will enable us to further
constrain the geometry of the system in Section 5. However, the fact
that inner eccentricity is not zero (as would be expected due to tidal
circularization of such a close binary) is evidence for interaction
with the third star. More convincingly, the inner and outer longitudes
are clearly varying. The roughly linear trend indicates a secular
advance of the line of apsides, a direct indication of the effect of
three-body interaction.

\subsection{Precession of the nodes}

\begin{figure*}
\def\epsfsize#1#2{0.95#1}
\epsfbox{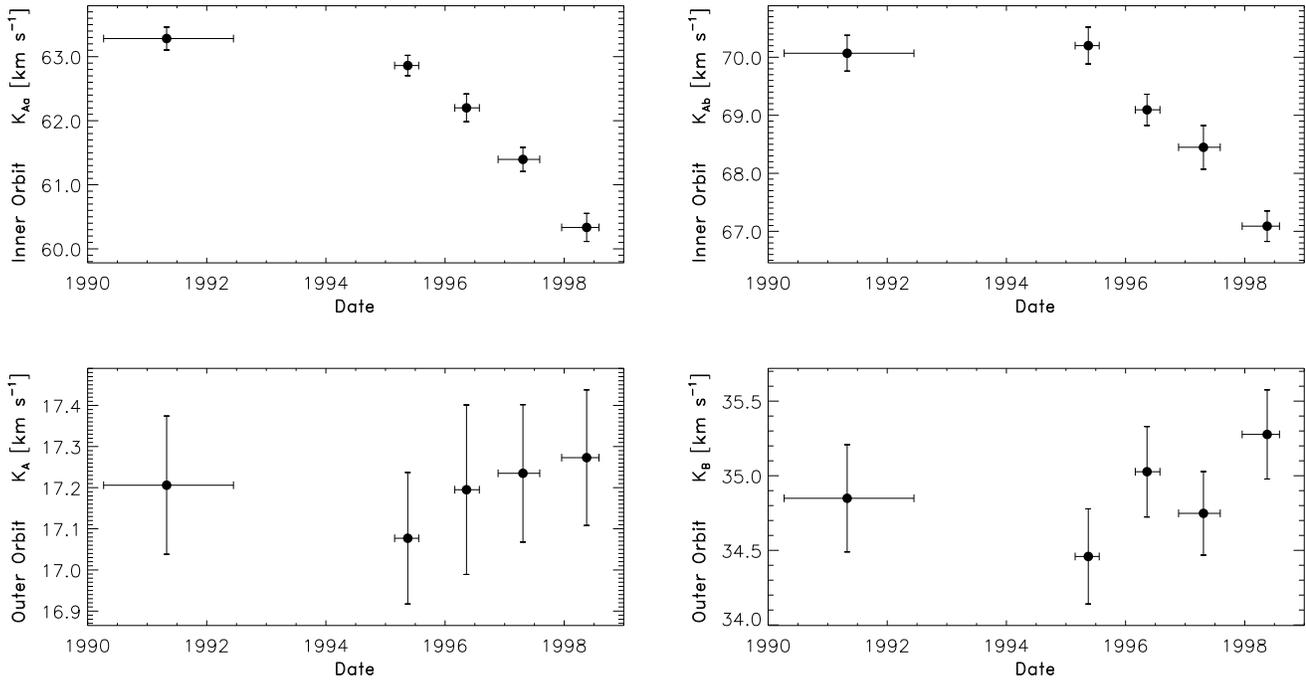}
\caption{Radial velocity amplitudes. The upper plots show $K_{Aa}$ and $K_{Ab}$
from the inner orbit, while the lower plots show $K_{A}$ and $K_{B}$ from the
outer orbit. The inner orbit shows clear evidence for the 
 precession of the nodes.}
\label{fig-kvar}
\end{figure*}

The radial velocity amplitudes, $K_{Aa}$, $K_{Ab}$, $K_A$ and $K_B$,
from the five subsets are shown in Figure \ref{fig-kvar}.
The amplitudes of the inner binary, $K_{Aa}$ and $K_{Ab}$, both show
very clear variation, with the same trend.
This is well understood if we assume the variation is caused by
the precession of the nodes. To show that this is the case we note the 
$K_{Aa} = V_{Aa} \sin i_A$, where $V$ is used to denote
the true orbital velocity amplitude rather than the projected radial
velocity amplitude, and $K_{Ab} = V_{Ab} \sin i_A$. Variation of the
inclination angle would thus induce the same trend for $K_{Aa}$ and $K_{Ab}$.

\begin{figure}
\def\epsfsize#1#2{0.95#1}
\epsfbox{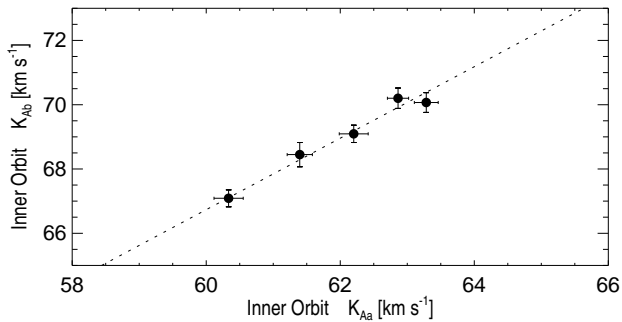}
\caption{Solutions from the five separate data subsets, plotted in
($K_{Aa}, K_{Ab}$) space. The dashed line is the best-fit straight
line which goes through the origin, expected for constant
$K_{Aa}/K_{Ab}$ ratio.}
\label{fig-kcons}
\end{figure}

Furthermore, if the variations of the observed amplitudes are caused
only by the modulation of the inclination angle, the ratio
$K_{Aa}/K_{Ab}$ should remain constant. To check this point we have
plotted in Figure \ref{fig-kcons} the results from the five subsets in
the ($K_{Aa}$, $K_{Ab}$) parameter space.  If the ratio between the
two amplitudes is constant, we expect these five points to fall on a
straight line that goes through the origin. The figure shows a beatiful
confirmation of this prediction.  The variation of the amplitude
combined with the results for the inner inclination given in Section 2
imply an approximately $4 \degr$ decrease in $i_A$ over the span of
the observations.

Were this a real effect due to precession of the nodes, we would
expect an increase in the outer inclination angle, and correspondingly
an increase in $K_A$ and $K_B$. We can calculate the amplitude of such
an effect by linearly approximating the decrease in the inner radial
velocity amplitudes. As above, we know that
\begin{equation}
\frac{\Delta K_{Aa}}{V_{Aa}} = \frac{\Delta K_{Ab}}{V_{Ab}} \quad
\hbox{and} \quad
\frac{\Delta K_{A}}{V_{A}} = \frac{\Delta K_{B}}{V_{B}}
\end{equation}
The lack of significant eccentricity modulation implies that the
relative inclination $\phi$ has remained nearly constant (because of
conservation of the total angular momentum), and validates equations
\ref{eqn-cosia} and \ref{eqn-cosiab}. Using these and differentiating
$K = V \sin i$ with respect to time yields
\begin{equation}
\frac{1}{V_{Aa}}\frac{dK_{Aa}}{dt} = \frac{1}{V_{Ab}}\frac{dK_{Ab}}{dt} = 
\omega_p \cot i_A \sin \alpha \sin \beta_A \sin [\omega_p (t - t_0)] \quad
\hbox{and}
\end{equation}
\begin{equation}
\frac{1}{V_{A}}\frac{dK_{A}}{dt} = \frac{1}{V_{B}}\frac{dK_{B}}{dt} =
- \omega_p \cot i_{AB} \sin \alpha \sin \beta_{AB} \sin [\omega_p (t - t_0)].
\end{equation}
Assuming that this first-derivative (i.e. linear) expansion is
sufficient to cover the span of our observations, we can take the
ratio of these two equations and determine that
\begin{equation}
\frac{\frac{\Delta K_{Ab}}{V_{Ab}}}{\frac{\Delta K_B}{V_B}} =
- \frac{\cot i_A}{\cot i_{AB}}\frac{\sin \beta_A}{\sin \beta_{AB}}.
\label{eqn-dk}
\end{equation}
Because $\bmath{G} = \bmath{G_A} + \bmath{G_{AB}}$, the law of sines
can be applied. For a binary orbit, the amplitude of the angular
momentum about the centre-of-mass is 
\begin{equation}
L \propto \mu [M a (1 -
e^2)]^{1/2} \propto \mu M^{2/3} P^{1/3} (1 - e^2)^{1/2},
\end{equation}
where $\mu$
is the reduced mass and $M$ is the total mass. Thus we have
\begin{equation}
\frac{\sin \beta_A}{\sin \beta_{AB}} =
\frac{\left|\bmath{G_{AB}}\right|}{\left|\bmath{G_{A}}\right|} =
\left(\frac{\mu_{AB}}{\mu_{A}}\right)
\left(\frac{M_{AB}}{M_{A}}\right)^{2/3}
\left(\frac{P_{AB}}{P_{A}}\right)^{1/3}
\left(\frac{1-e_{AB}^2}{1-e_{A}^2}\right)^{1/2}
\label{eqn-sinbeta}
\end{equation}
Figure \ref{fig-kvar} shows that $\Delta K_{Ab} \approx -2.9 \;
\rmn{km~s^{-1}}$. Combining equations \ref{eqn-dk} and
\ref{eqn-sinbeta} and inserting the parameters for HD 109648 from
Section 2, we expect $\Delta K_{B} \approx 0.4 \;
\rmn{km~s^{-1}}$. Our error bars are too large to claim a detection of
this, but the expectation is consistent with what we
observe. Continued observations may make the precession more clear.

The already observed precession also enables us to strengthen our
lower limit on the relative inclination, $\phi_{\rm min}$, calculated
in Section 2. Because $i_A$ has been decreasing significantly, while
$i_{AB}$ has likely increased slightly, the difference $i_A-i_{AB}$,
(which is a strict lower limit on the relative inclination) has also
been decreasing. Thus our strongest constraint on $\phi_{\rm min}$ can
come from analyzing our earliest subset of data only, where
$i_A-i_{AB}$ was the greatest, rather than its average value over the
whole time span. This yields a refined minimum relative inclination,
$\phi_{\rm min} = 5.4 \pm 0.4 \degr$.

\section{Simulation}

The amplitude of the inner eccentricity modulation is especially
sensitive to the relative inclination, $\phi$, between the two orbital
planes (Mazeh \& Shaham 1979; Bailyn 1987; Paper II). The
modulation amplitude increases with the relative inclination. For
relative inclinations greater than a critical relative inclination,
$\phi \ga \phi_{\rmn{crit}} \sim 40 \degr$, the modulation amplitude
increases dramatically, with the possibility of the inner eccentricity
approaching unity. Our observations, spanning more than eight years, have
yielded an average inner eccentricity of $e_A = 0.0119 \pm
0.0014$. Using numerical simulations, we can estimate the likelihood of
this result for different values of the relative inclination angle.

Our simulations are similar to those described in Paper I, integrating
Newton's equations for three mass points. The starting point for the
integrations was determined from the elements over all the
observations, given in Table \ref{tab-elem}. We have used the three-body
regularization program of Aarseth \cite{aar74}, as well as a code
written by Bailyn \shortcite{bai87}, to perform the integrations.  We
have also developed and used a code written specifically for this
system. All three routines yielded identical results.

\begin{figure*}
\def\epsfsize#1#2{0.95#1}
\epsfbox{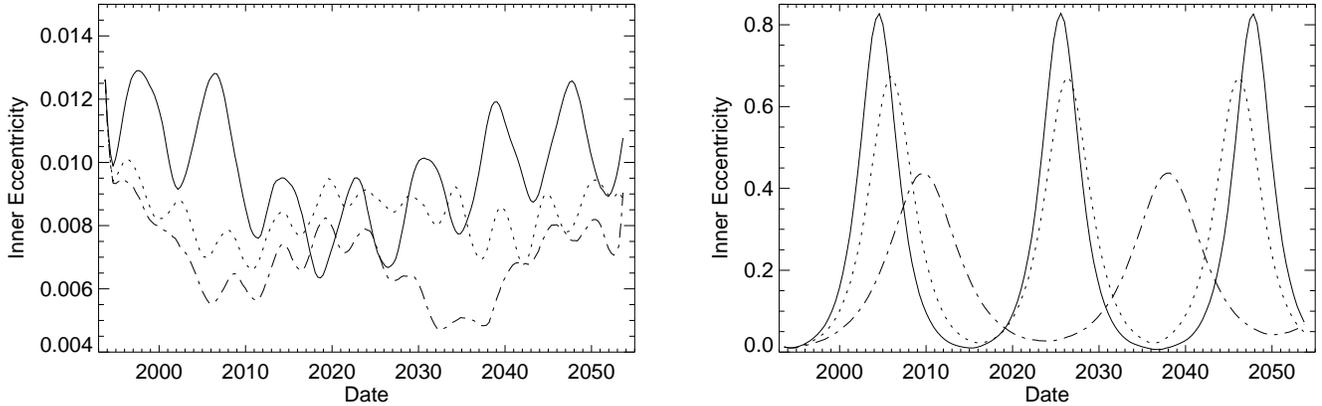}
\caption{Typical simulation results for the inner eccentricity modulation. The
left panel shows the modulation for low relative inclination (dash-dotted:
$\phi = 6\degr$, dotted: $\phi = 15\degr$, solid: $\phi = 30\degr$). The
modulation shown in the right panel is for high relative inclination
(dash-dotted: $\phi = 50\degr$, dotted: $\phi = 60\degr$, solid: $\phi =
70\degr$).}
\label{fig-sim}
\end{figure*}

Typical results of the simulations are shown in Figure
\ref{fig-sim}. As discussed in Paper II, the eccentricity modulation
depends not only on the relative inclination, but also on the
arguments of periastron, $g_A$ and $g_{AB}$, measured with respect to
the unknown intersection of the orbital planes (S\"{o}derhjelm 1984;
Paper II). These arguments are especially important at the lower
relative inclinations. We have explored a range of values, so that our
simulations are typical cases.

As the figure shows, for small relative inclinations an eight-year window
could easily produce an eccentricity modulation consistent with what we have
observed.  At higher relative inclinations, however, the probability of
obtaining a small average inner eccentricity over eight years decreases. By
running many different simulations, we can quantify this probability, and
estimate an upper limit to the relative inclination, $\phi_{\rmn{max}} \simeq
54 \degr$ above which a low inner eccentricity cannot be maintained for eight
years. However, our simulations were limited, considering only Newtonian
gravity with three point masses. Other effects, including quadrupole
perturbations in the inner binary, tidal friction and general relativistic
effects, may be significant factors in the eccentricity modulation. In
particular, such effects may dampen the eccentricity modulation amplitude at
high relative inclinations and thus our estimate of the upper limit on the
relative inclination is not very firm.

On the other hand, we have a strong lower limit on the relative inclination
calculated in Section 2 from the geometry of the system and strengthened in
Section 4 via the precession of the nodes. Using this technique, thus, we can
significantly constrain the relative inclination. It turns out that the
ambiguity in the inclination angles corresponds to an outer orbit which either
co-rotates or counter-rotates with respect to the inner orbit. Choosing the
inclination angles in the same quadrant corresponds to the co-rotational case,
with limits on the relative inclination, \begin{equation} 5.4 \degr \leq \phi
\la 54 \degr \quad \hbox{(co-rotation).}
\end{equation}
The counter-rotational case leads to alternative limits for the relative
inclination,
 \begin{equation}
126 \degr \la \phi \leq 174.6 \degr \quad \hbox{(counter-rotation).}
\end{equation}
The relative inclinations closer to $\phi_{\rmn{max}}$ (i.e. the upper limit
in the co-rotational case or the lower limit in the counter-rotational case)
are less probable than those closer to $\phi_{\rmn{min}}$.

\section{Discussion}

We have shown that HD 109648 is a hierarchical triple system, with an
outer period and outer:inner period ratio conducive to modulations of
orbital elements on timescales of about a decade. Indeed, our
observations clearly indicate an advance of the inner longitude of
periastron corresponding roughly to this timescale. We also found
strong evidence for variations of the radial velocity amplitudes of
the inner orbit, most naturaly accounted for by the precession of the
nodes.  Furthermore, the inner eccentricity is small but significant,
presumably due to the interaction with the outer star.

Such effects have been predicted theoretically for hierarchical
triples for a number of years. However, there have been few
observational confirmations.  Mayor and Mazeh \shortcite{may87} have
looked for evidence of the precession of nodes in a number of close
binaries, and have reported several significant changes, based on
observations made at two widely-spaced epochs. Mazeh and Shaham
\shortcite{maz76} also suggested a few systems where the effect may
have had a role, but none of these were confirmed triples.

The inner eccentricity modulation has been less conclusively observed. Mazeh
and Shaham (1977; 1979) have postulated it to be the cause of long-period
phenomena, such as episodic accretion, in some close binaries. More convincing
evidence has been offered for the interaction of a third star with the tidal
circularization of the inner binary. Mazeh \shortcite{maz90} has looked for
eccentric orbits in samples of short-period binaries that should have been
circularized, as a fingerprint for a third star in the system. Three such
examples were found, and the hypothesis of a triple system was confirmed in
each case. In addition, one system (HD 144515) showed evidence for a variation
in the inner eccentricity, again based on observations from two epochs. Ford,
Kozinsky, and Rasio (1999) also provide instances of some other triple systems
where these interactions may have played a role.

HD 109648 provides the best observational evidence so far of these
predicted modulations. This system is a confirmed hierarchical triple,
a direct result of analysing the triple-lined spectra. Furthermore,
the evidence for variations in the elements comes from observations in
a homogeneous set of data, rather than relying on two-epoch
observations. Finally, we see evidence both for the precession of the
nodes and for the apsidal advance in the same system.

More data are needed to strengthen this case. With better information
on the variation of the elements with time, we should be able to
derive better constraints on the orientation of the system.  For
example, if we are able to fit the variation of the inner and outer
inclination angles, we can determine the various angles between the
total angular momentum and its inner and outer binary components, as
well as the angle between the total angular momentum and the line of
sight.  If the inner eccentricity modulation becomes clearer with
additional data as well, it should provide stronger constraints on the
relative inclination.

In addition to continued spectroscopic observations, there may be some
hope that interferometric observations of HD 109648 will be able to
help clarify the orientation of the system. The Hipparcos parallax for
HD 109648 of 4 mas may potentially be in error due to the outer orbit,
particularly because the outer period is nearly commensurate with one
year. Nevertheless, the separation of the third star from the inner
binary is on the order of a few mas, allowing for the possibility of a
visual orbit in the future.

We can also hope that additional triple systems will be discovered,
perhaps ones that are even better than HD 109648 for this type of
study. Indeed, Saar, N\"{o}rdstrom \& Andersen \shortcite{saa90} have
noted a promising system with a modulation timescale perhaps shorter
even than that of HD 109648. Determining the geometry and orientation
of such systems will be a great advance in our understanding of them.
 
\section*{Acknowledgments} 
 
We would like to thank J.~Caruso and J.~Zajac for making many of the
observations presented here, as well as R.~Davis for help with the data
reduction. We are grateful for the efforts of S. Chanmugam, T. Lynn, M. Morgan
and E. Scanapieco for their initial work in determining periods for the two
orbits. Thanks also go to S.~Aarseth and C.~Bailyn for use of their three-body
codes. We thank the anonymous referee for providing useful suggestions. This
work was supported by US-Israel Binational Science Foundation grant 94-00284
and 97-00460 and an NSF Graduate Research Fellowship. SJ thanks the Harvard
University Department of Astronomy for its support of this work.  SJ would
also expresses his sincere gratitude to Y.~Krymolowski for many helpful
discussions and his wonderful generosity.

\appendix 
 
\section{Radial velocities} 
In the following table we list our observations of the heliocentric 
radial velocities for the three visible components of HD109648. The 
date given is HJD - 2400000, and the velocities are in 
$\rmn{km~s^{-1}}$. 
 
\begin{table} 
\label{tab-vel} 
\caption{HD109648 radial velocities} 
\begin{tabular}{crrr} 
Date & $v_{Aa}$ & $v_{Ab}$ & $v_{B}$ \\ 
\hline 
47989.7237 &     8.27 &   -71.69 &     3.49 \\  
47999.7229 &   -61.78 &    -0.48 &     5.75 \\  
48020.6318 &   -91.61 &    38.78 &     5.81 \\  
48026.5976 &   -81.05 &    35.39 &    -5.62 \\  
48047.6212 &   -56.23 &    58.64 &   -55.31 \\  
48050.6484 &    62.67 &   -70.83 &   -54.71 \\  
48058.5622 &   -53.61 &    62.15 &   -59.52 \\  
48059.5741 &   -50.88 &    56.89 &   -56.80 \\  
48078.5968 &    43.71 &   -75.39 &   -33.52 \\  
48079.5639 &   -14.67 &   -14.66 &   -31.02 \\  
48082.5545 &     2.21 &   -32.98 &   -20.68 \\  
48083.5582 &    46.24 &   -86.64 &   -23.75 \\  
48084.5680 &    13.92 &   -49.12 &   -21.77 \\  
48086.5908 &   -81.22 &    47.77 &   -20.62 \\  
48087.5809 &   -36.27 &    -3.84 &   -23.73 \\  
48088.5697 &    30.87 &   -77.91 &   -17.47 \\  
48089.5811 &    34.05 &   -83.29 &   -16.95 \\  
48102.5641 &   -88.51 &    38.11 &    -3.03 \\  
48108.5458 &   -90.10 &    38.41 &    -0.47 \\  
48226.8598 &    11.62 &   -73.25 &     1.26 \\  
48251.9081 &   -47.72 &   -11.73 &    10.72 \\  
48279.9641 &    17.77 &   -52.29 &   -32.17 \\  
48288.8245 &   -61.96 &    63.08 &   -52.66 \\  
48309.7793 &   -15.70 &     3.53 &   -46.37 \\  
48327.8502 &   -75.92 &    43.35 &   -19.55 \\  
48349.7883 &   -86.76 &    32.36 &     1.96 \\  
48354.7527 &   -94.22 &    38.24 &     5.32 \\  
48372.6173 &   -35.18 &   -30.36 &    10.47 \\  
48380.6620 &   -16.25 &   -42.88 &     6.52 \\  
48385.6606 &    15.24 &   -77.57 &    -2.72 \\  
48395.6641 &    40.76 &   -84.38 &   -19.89 \\  
48409.6348 &   -61.88 &    65.74 &   -57.33 \\  
48413.6430 &     7.05 &    -5.73 &   -58.60 \\  
48414.6856 &   -53.57 &    61.53 &   -58.13 \\  
48415.6193 &   -50.53 &    61.78 &   -58.18 \\  
48428.5738 &    57.32 &   -73.48 &   -49.80 \\  
48429.5644 &    35.66 &   -52.56 &   -47.59 \\  
48431.5744 &   -67.68 &    57.96 &   -46.16 \\  
48432.5571 &   -31.54 &    24.11 &   -43.70 \\  
48433.5852 &    36.34 &   -56.62 &   -39.55 \\  
48434.5791 &    52.37 &   -75.33 &   -41.49 \\  
48435.5893 &    -4.54 &   -15.46 &   -35.88 \\  
48438.5786 &    -1.85 &   -26.92 &   -34.98 \\  
48439.5541 &    48.44 &   -81.45 &   -27.24 \\  
48442.5782 &   -76.91 &    52.03 &   -26.14 \\  
48454.5700 &   -41.14 &    -3.22 &   -11.80 \\  
48456.5764 &    34.51 &   -88.69 &   -14.46 \\  
48458.5707 &   -82.10 &    39.30 &    -7.50 \\  
48462.5412 &     9.60 &   -64.66 &    -4.10 \\  
48466.5641 &    19.99 &   -81.09 &    -0.80 \\  
48469.5647 &   -86.54 &    33.22 &     1.36 \\  
48601.9412 &   -82.25 &    26.99 &     9.14 \\  
48612.8884 &   -83.48 &    24.34 &     9.55 \\  
48621.9574 &   -34.13 &   -24.48 &     3.54 \\  
48632.8870 &   -24.36 &   -24.90 &    -9.53 \\  
48638.8574 &   -51.46 &    22.95 &   -23.83 \\  
48662.8834 &   -10.31 &    11.97 &   -58.74 \\  
48665.8606 &   -10.21 &     9.42 &   -56.13 \\  
48669.9563 &    59.43 &   -74.15 &   -50.67 \\  
48679.7531 &     7.64 &   -36.26 &   -33.01 \\  
48691.7352 &    44.20 &   -91.01 &   -14.88 \\  
\end{tabular} 
\end{table}

\begin{table} 
\contcaption{} 
\begin{tabular}{crrr} 
Date & $v_{Aa}$ & $v_{Ab}$ & $v_{B}$ \\ 
\hline 
48706.7376 &   -35.67 &   -15.70 &     0.75 \\  
48711.7672 &   -66.89 &    13.10 &     2.95 \\  
48721.6201 &   -93.70 &    34.41 &    10.33 \\  
48726.6840 &   -78.76 &    16.65 &     7.98 \\  
48759.6447 &   -65.65 &    36.28 &   -25.29 \\  
48768.6163 &    58.58 &   -74.76 &   -47.66 \\  
48785.6718 &    40.20 &   -45.64 &   -58.69 \\  
49045.8374 &   -60.86 &    34.77 &   -27.26 \\  
49065.8097 &   -38.38 &   -17.99 &    -3.76 \\  
49082.7215 &   -72.76 &    13.25 &     7.67 \\  
49092.6583 &    -5.49 &   -61.83 &     4.93 \\  
49106.6988 &   -34.95 &   -24.62 &     1.48 \\  
49115.7127 &   -69.84 &    28.48 &   -12.82 \\  
49773.7397 &   -80.37 &    50.52 &   -19.63 \\  
49786.8017 &    17.02 &   -77.36 &    -5.51 \\  
49795.7514 &   -90.82 &    41.62 &     2.87 \\  
49796.7069 &   -53.68 &    -2.26 &     3.79 \\  
49799.7128 &   -26.93 &   -32.54 &     5.53 \\  
49801.8641 &   -73.75 &    18.64 &     8.16 \\  
49802.7256 &   -17.64 &   -47.75 &     6.78 \\  
49812.7374 &   -79.52 &    19.43 &    10.31 \\  
49813.6946 &   -18.40 &   -49.16 &     9.11 \\  
49814.8387 &    30.79 &  -102.61 &     8.91 \\  
49815.7253 &     0.69 &   -69.38 &     7.59 \\  
49817.7980 &   -92.81 &    34.62 &     8.20 \\  
49818.6922 &   -51.15 &    -9.88 &     9.38 \\  
49819.7020 &    14.21 &   -86.23 &     7.75 \\  
49822.7849 &   -91.01 &    35.77 &     7.31 \\  
49824.6775 &   -13.81 &   -48.19 &     5.84 \\  
49825.8203 &    33.53 &  -101.22 &     5.07 \\  
49828.6516 &   -90.42 &    39.92 &     3.55 \\  
49831.6661 &    30.23 &   -91.54 &     0.11 \\  
49832.6810 &   -28.65 &   -23.83 &    -2.44 \\  
49844.6766 &   -74.82 &    49.36 &   -25.82 \\  
49845.6847 &   -57.39 &    31.17 &   -25.54 \\  
49846.6568 &     7.51 &   -36.18 &   -30.87 \\  
49847.6814 &    52.54 &   -82.92 &   -31.62 \\  
49851.6380 &   -20.41 &     5.17 &   -42.99 \\  
49854.6451 &   -11.74 &     0.60 &   -48.20 \\  
49857.6233 &    21.91 &   -26.35 &   -52.17 \\  
49858.6378 &    63.86 &   -70.39 &   -54.37 \\  
49860.6530 &   -41.31 &    45.52 &   -59.99 \\  
49861.5812 &   -60.33 &    73.00 &   -58.02 \\  
49873.5634 &   -16.18 &    13.82 &   -50.70 \\  
49874.6595 &    50.08 &   -64.86 &   -51.72 \\  
49875.6122 &    50.61 &   -64.67 &   -48.90 \\  
49876.5639 &   -10.27 &     3.28 &   -48.88 \\  
49884.6025 &   -19.25 &    -2.15 &   -33.72 \\  
49885.6143 &    42.48 &   -70.01 &   -31.00 \\  
49886.5958 &    40.49 &   -69.83 &   -29.20 \\  
49887.5696 &   -23.48 &    -1.81 &   -25.58 \\  
49888.5820 &   -73.23 &    53.88 &   -27.01 \\  
49889.5828 &   -54.25 &    30.32 &   -23.22 \\  
49890.6078 &    12.64 &   -47.17 &   -22.63 \\  
49891.5853 &    47.28 &   -85.71 &   -20.76 \\  
49901.5663 &     5.02 &   -53.63 &    -9.04 \\  
49903.5623 &    -1.71 &   -47.24 &    -6.70 \\  
49905.5629 &   -82.56 &    39.32 &    -5.78 \\  
\end{tabular} 
\end{table}

\begin{table} 
\contcaption{} 
\begin{tabular}{crrr} 
Date & $v_{Aa}$ & $v_{Ab}$ & $v_{B}$ \\ 
\hline 
49909.5579 &   -42.72 &   -11.48 &    -3.03 \\  
49911.5591 &   -63.54 &    15.05 &     1.49 \\  
49913.6108 &    34.38 &   -97.08 &     3.70 \\  
49918.5850 &    27.62 &   -94.18 &     4.70 \\  
49920.5635 &   -50.20 &    -8.34 &     8.08 \\  
49921.5781 &   -92.65 &    36.67 &     7.68 \\  
50063.9522 &   -91.42 &    36.99 &     7.72 \\  
50142.7836 &    14.33 &   -66.60 &    -9.83 \\  
50143.8448 &    35.45 &   -90.43 &    -8.06 \\  
50146.7695 &   -78.21 &    33.86 &    -3.59 \\  
50151.8049 &   -90.40 &    39.96 &    -1.16 \\  
50152.8799 &   -46.78 &    -6.37 &     0.80 \\  
50153.7820 &    12.21 &   -74.33 &     3.74 \\  
50154.8156 &    29.80 &   -95.35 &     1.76 \\  
50155.8299 &   -26.69 &   -30.71 &     2.78 \\  
50156.6971 &   -80.02 &    26.01 &     3.80 \\  
50172.7725 &   -64.97 &     4.87 &     7.90 \\  
50173.8092 &   -93.74 &    34.74 &     9.02 \\  
50174.8187 &   -48.98 &   -12.96 &     9.56 \\  
50176.8073 &    24.37 &   -96.12 &     8.06 \\  
50177.7284 &   -30.72 &   -33.34 &     7.19 \\  
50185.7026 &   -51.35 &    -6.47 &     5.49 \\  
50198.5911 &    35.74 &   -90.65 &    -8.14 \\  
50201.6261 &   -69.71 &    33.64 &   -14.13 \\  
50202.6803 &    -3.42 &   -37.85 &   -15.35 \\  
50206.6064 &   -76.90 &    52.93 &   -25.11 \\  
50210.6036 &    -9.49 &   -10.95 &   -35.23 \\  
50211.7135 &   -67.17 &    56.52 &   -39.18 \\  
50216.6774 &   -40.16 &    38.35 &   -48.67 \\  
50228.7005 &   -57.98 &    68.09 &   -60.26 \\  
50229.5814 &   -17.83 &    21.07 &   -61.32 \\  
50233.5826 &   -58.14 &    62.18 &   -53.43 \\  
50235.6004 &    16.07 &   -21.89 &   -51.92 \\  
50236.6059 &    59.03 &   -69.41 &   -50.02 \\  
50237.6100 &    27.84 &   -37.62 &   -49.40 \\  
50239.6119 &   -65.37 &    62.10 &   -44.99 \\  
50240.6300 &   -19.83 &     8.50 &   -45.50 \\  
50242.5957 &    47.45 &   -69.39 &   -40.01 \\  
50258.5729 &    42.22 &   -89.86 &   -15.70 \\  
50260.5793 &   -61.65 &    21.44 &   -12.95 \\  
50261.5914 &   -81.30 &    42.94 &   -11.88 \\  
50263.5636 &    26.28 &   -79.50 &    -9.25 \\  
50266.6003 &   -85.36 &    42.25 &    -6.34 \\  
50270.5630 &     1.03 &   -55.75 &    -2.40 \\  
50275.5685 &    22.59 &   -84.41 &     1.81 \\  
50285.5636 &    23.13 &   -92.08 &     7.40 \\  
50287.5558 &   -45.50 &   -14.82 &     9.06 \\  
50292.5516 &   -11.93 &   -55.50 &    11.11 \\  
50293.5451 &   -77.62 &    19.16 &    10.68 \\  
50407.9227 &   -31.85 &   -31.77 &     7.18 \\  
50415.9494 &   -43.78 &   -20.20 &    10.36 \\  
50437.9290 &   -29.53 &   -11.90 &    -5.62 \\  
50443.9209 &     6.92 &   -52.32 &   -18.57 \\  
50460.9211 &    54.87 &   -62.26 &   -57.42 \\  
50462.9635 &   -20.03 &    21.95 &   -58.43 \\  
50465.8867 &    32.72 &   -30.65 &   -60.69 \\  
50467.9641 &    15.05 &   -14.34 &   -62.34 \\  
\end{tabular} 
\end{table} 
 
\begin{table} 
\contcaption{} 
\begin{tabular}{crrr} 
Date & $v_{Aa}$ & $v_{Ab}$ & $v_{B}$ \\ 
\hline 
50472.9241 &    45.63 &   -54.75 &   -56.50 \\  
50477.8895 &    56.86 &   -69.28 &   -51.02 \\  
50481.9188 &    -5.94 &    -6.37 &   -44.48 \\  
50492.8095 &   -19.26 &   -15.75 &   -22.29 \\  
50495.8587 &   -39.35 &     7.02 &   -18.69 \\  
50499.8407 &    40.18 &   -90.05 &   -14.82 \\  
50502.8649 &   -74.78 &    36.56 &   -10.20 \\  
50504.8512 &    33.71 &   -93.86 &    -9.28 \\  
50507.8283 &   -85.63 &    42.91 &    -3.88 \\  
50514.8631 &   -19.30 &   -38.91 &     5.08 \\  
50516.8764 &    11.82 &   -71.16 &     0.87 \\  
50521.7889 &    28.84 &   -98.68 &     3.24 \\  
50523.7647 &   -83.61 &    21.19 &     2.86 \\  
50526.6975 &    24.94 &   -92.82 &     6.81 \\  
50528.8336 &   -61.17 &     2.61 &     6.38 \\  
50530.8489 &   -54.04 &    -7.41 &     8.86 \\  
50535.6834 &   -87.28 &    28.47 &     9.36 \\  
50540.7280 &   -93.20 &    35.94 &     7.23 \\  
50543.7510 &    27.71 &   -98.32 &     7.76 \\  
50548.7202 &    29.56 &   -94.34 &     2.31 \\  
50550.7073 &   -56.17 &    -1.10 &     5.47 \\  
50552.7438 &   -49.91 &    -7.01 &     1.65 \\  
50554.6823 &    32.37 &   -93.86 &    -1.04 \\  
50559.7188 &    35.81 &   -91.00 &    -8.35 \\  
50561.6376 &   -47.89 &     5.79 &   -13.11 \\  
50565.6876 &    39.91 &   -81.27 &   -20.57 \\  
50567.6871 &   -71.29 &    49.65 &   -22.03 \\  
50569.6653 &    -2.94 &   -27.34 &   -27.64 \\  
50569.7151 &    -0.35 &   -27.81 &   -28.61 \\  
50571.6093 &    29.76 &   -56.43 &   -32.20 \\  
50573.5884 &   -69.39 &    58.27 &   -39.81 \\  
50581.6726 &    60.56 &   -66.47 &   -55.04 \\  
50583.6745 &   -32.28 &    37.28 &   -59.00 \\  
50584.5720 &   -59.02 &    70.28 &   -60.46 \\  
50586.6867 &    46.14 &   -46.93 &   -61.62 \\  
50587.5819 &    62.05 &   -64.78 &   -61.81 \\  
50589.5764 &   -50.38 &    62.36 &   -60.94 \\  
50591.6532 &    15.57 &   -15.47 &   -61.18 \\  
50595.6468 &   -60.57 &    67.52 &   -57.21 \\  
50597.5898 &    39.31 &   -55.07 &   -48.82 \\  
50608.5644 &    31.64 &   -56.79 &   -31.49 \\  
50610.5844 &   -13.25 &   -13.34 &   -29.83 \\  
50614.5538 &    42.61 &   -82.71 &   -22.75 \\  
50616.6050 &   -53.81 &    21.35 &   -19.12 \\  
50619.5932 &    26.68 &   -70.98 &   -14.71 \\  
50622.6245 &   -80.05 &    40.79 &   -12.74 \\  
50624.6218 &    -1.81 &   -48.60 &   -10.44 \\  
50627.5643 &   -61.05 &    13.59 &    -6.69 \\  
50629.5753 &   -41.50 &   -10.38 &    -2.73 \\  
50631.6240 &    24.56 &   -87.12 &    -0.82 \\  
50634.5747 &   -71.58 &    14.22 &    -1.88 \\  
50636.5764 &    32.29 &   -97.93 &     2.85 \\  
50640.5665 &   -41.46 &   -15.53 &     4.26 \\  
50642.5644 &    21.57 &   -90.07 &     4.42 \\  
50646.5856 &    -8.72 &   -58.15 &     5.18 \\  
50649.5672 &   -74.46 &    12.44 &     6.12 \\  
50653.5645 &    19.97 &   -90.01 &     8.11 \\  
50655.5699 &   -91.39 &    32.95 &     8.40 \\  
50656.5551 &   -72.95 &    14.22 &     5.63 \\  
50660.5468 &   -74.84 &    15.02 &     1.76 \\
50663.5513 &    21.44 &   -89.77 &     8.56 \\
\end{tabular} 
\end{table} 
 
\begin{table} 
\contcaption{} 
\begin{tabular}{crrr} 
Date & $v_{Aa}$ & $v_{Ab}$ & $v_{B}$ \\ 
\hline 
50797.9062 &   -83.63 &    37.29 &    -5.48 \\
50803.8913 &   -79.47 &    43.31 &   -12.35 \\
50814.9400 &   -66.79 &    53.63 &   -41.84 \\
50825.9076 &   -55.69 &    64.16 &   -60.69 \\
50834.9095 &    22.09 &   -23.55 &   -58.83 \\
50910.6468 &    29.72 &   -96.99 &     4.00 \\
50911.6455 &   -11.34 &   -49.39 &     2.67 \\
50914.6172 &   -29.06 &   -27.43 &     0.81 \\
50915.6724 &    26.84 &   -90.62 &     0.58 \\
50917.7045 &   -48.18 &     0.08 &    -3.80 \\
50918.7243 &   -86.02 &    40.70 &    -3.73 \\
50919.7225 &   -51.31 &     4.70 &    -4.78 \\
50922.6137 &    -5.00 &   -43.91 &   -11.45 \\
50924.6904 &   -71.70 &    36.36 &   -14.91 \\
50925.6614 &   -17.19 &   -19.51 &   -15.24 \\
50929.6813 &   -76.37 &    51.08 &   -25.92 \\
50931.6316 &    21.59 &   -51.75 &   -30.75 \\
50932.6945 &    44.88 &   -76.60 &   -32.51 \\
50934.6477 &   -59.04 &    46.01 &   -37.73 \\
50945.7289 &   -49.81 &    62.03 &   -58.25 \\
50946.7063 &   -46.23 &    55.96 &   -59.89 \\
50948.6479 &    59.44 &   -64.30 &   -59.96 \\
50949.6324 &    42.59 &   -44.57 &   -60.96 \\
50950.6296 &   -22.94 &    29.47 &   -63.94 \\
50951.7112 &   -61.31 &    68.11 &   -57.60 \\
50952.7105 &   -18.26 &    23.38 &   -60.51 \\
50953.5900 &    36.55 &   -37.49 &   -61.43 \\
50954.6430 &    58.71 &   -64.29 &   -58.83 \\
50956.5913 &   -50.20 &    54.90 &   -55.59 \\
50958.6129 &     5.79 &   -10.27 &   -52.67 \\
50961.6626 &   -33.40 &    27.84 &   -49.03 \\
50964.5528 &    28.95 &   -49.47 &   -42.92 \\
50966.5663 &     1.02 &   -21.17 &   -39.88 \\
50968.5815 &   -60.31 &    44.96 &   -33.71 \\
50974.5877 &   -39.63 &     7.86 &   -28.00 \\
50976.5979 &    40.66 &   -80.78 &   -20.83 \\
50993.5849 &     6.59 &   -66.82 &     0.10 \\
50996.6379 &   -41.89 &   -12.75 &     0.72 \\
50997.5898 &    14.54 &   -79.85 &    -0.51 \\
50998.5910 &    25.37 &   -91.13 &     3.73 \\
51000.5831 &   -86.95 &    27.19 &     4.43 \\
51000.6085 &   -86.04 &    31.18 &     2.89 \\
51004.5778 &     0.20 &   -64.65 &     4.94 \\
51005.5690 &   -65.26 &     7.19 &     3.71 \\
51006.5708 &   -90.43 &    34.43 &     5.64 \\
51008.5673 &    15.48 &   -83.70 &    10.36 \\
51009.5967 &    25.15 &   -90.61 &    13.22 \\
51013.5759 &   -12.46 &   -54.37 &     9.80 \\
51014.5698 &    26.17 &   -99.64 &     7.41 \\
51019.5842 &    15.79 &   -87.62 &     9.55 \\
51020.5530 &    21.55 &   -92.12 &     8.06 \\
51024.5518 &   -11.33 &   -56.40 &     7.86 \\
51026.5814 &    -5.79 &   -60.56 &     5.04 \\
\end{tabular} 
\label{lastpage} 
\end{table} 
         

\begin{thebibliography}{} 
\bibitem[\protect\citename{Aarseth \& Zare }1974]{aar74}Aarseth S. J., 
Zare K., 1974, Celest. Mech., 10, 185 
\bibitem[\protect\citename{Bailyn }1987]{bai87}Bailyn C. D., 1987, 
Ph.D. thesis, Harvard University 
\bibitem[\protect\citename{Batten }1973]{bat73}Batten A. H., 1973, 
Binary and Multiple Systems of Stars, Pergamon Press, New York 
\bibitem[\protect\citename{Baumgardt }1998]{bau98}Baumgardt, H. 1998,
A\&A, 340, 402
\bibitem[\protect\citename{Duquennoy \& Mayor }1991]{duq91}Duquennoy 
A., Mayor M., 1991, A\&A, 248, 485 
\bibitem[\protect\citename{Fekel }1981]{fek81}Fekel F. C., Jr, 1981, 
ApJ, 246, 879 
\bibitem[\protect\citename{Ford et al. }1999]{for99}Ford E. B., Kozinsky B.,
Rasio F. A., 1999, ApJ, in press (astro-ph/9905348)
\bibitem[\protect\citename{Gatewood et al. }1988]{gat88}Gatewood G., De Jonge 
J. K., Castelaz M., Han I., Persinger T., Prosser J., Reiland T., 
Stein J., 1988, ApJ, 332, 917 
\bibitem[\protect\citename{Gray }1992]{gra92}Gray, D. F., 1992, The 
Observation and Analysis of Stellar Photospheres, Cambridge University
Press, Cambridge
\bibitem[\protect\citename{Griffin }1992]{gri92}Griffin R. F., 1992, 
in Duquennoy A., Mayor M., eds, Binaries as Tracers of Stellar 
Formation. Cambridge Univ. Press, Cambridge, p. 96 
\bibitem[\protect\citename{Hale }1994]{hal94}Hale, A., 1994, AJ, 107, 306 
\bibitem[\protect\citename{Holman, Touma \& Tremaine
}1997]{hol97}Holman M., Touma J., Tremaine S., 1997, Nature, 386, 254
\bibitem[\protect\citename{Horne \& Baliunas }1986]{hor86}Horne J. H., 
Baliunas S. L., 1986, ApJ, 302, 757 
\bibitem[\protect\citename{Jha et al. }1997]{jha97}Jha S., Torres G., 
Stefanik R. P., Latham D. W., 1997, Baltic Astronomy, 6, 55 
\bibitem[\protect\citename{Krymolowski }1995]{kry95}Krymolowski Y., 
1995, M.Sc. thesis, Tel Aviv University 
\bibitem[\protect\citename{Paper II}]{pap2}Krymolowski Y., Mazeh T.,
1998, MNRAS, 304, 720 (Paper II)
\bibitem[\protect\citename{Kurucz }1992]{kur92}Kurucz R. L., 1992, 
Proc. IAU Symp. 149, 225 
\bibitem[\protect\citename{Latham }1985]{lat85}Latham D. W., 1985, in
Philip A. G. D., Latham D. W., eds, Stellar Radial Velocities, IAU
Coll. No. 88, L. Davis Press, Schnectady, p. 21
\bibitem[\protect\citename{Latham et al. }1988]{lat88}Latham D. W., 
Mazeh T., Carney B. W., McCrosky R. E., Stefanik R. P., Davis R. J., 
1988, AJ, 96, 567
\bibitem[\protect\citename{Latham }1992]{lat92}Latham D. W., 1992, in 
McAlister H. A., Hartkopf W. I., eds, Complementary Approaches to 
Double and Multiple Star Research, IAU Colloq. 135, ASP Conference 
Series v. 32, p. 110 
\bibitem[\protect\citename{Mathieu \& Mazeh }1988]{mat88}Mathieu,
R.D., Mazeh, T., 1988, ApJ, 326, 256
 \bibitem[\protect\citename{Mathieu et al. }1992]{mat92}Mathieu R. D., 
Duquennoy A., Latham D. W., Mayor M., Mazeh T., Mermilliod, J.-C., 
1992, in Duquennoy A., Mayor M., eds, Binaries as Tracers of Stellar 
Formation, Cambridge Univ. Press, Cambridge, p. 278 
\bibitem[\protect\citename{Mayor \& Mazeh }1987]{may87}Mayor M., Mazeh 
T., 1987, A\&A, 171, 157 
\bibitem[\protect\citename{Mazeh }1990]{maz90}Mazeh T., 1990, AJ, 99, 675 
\bibitem[\protect\citename{Mazeh \& Shaham }1976]{maz76}Mazeh T., 
Shaham J., 1976, ApJ, 208, L147 
\bibitem[\protect\citename{Mazeh \& Shaham }1977]{maz77}Mazeh T., 
Shaham J., 1977, ApJ, 213, L17 
\bibitem[\protect\citename{Mazeh \& Shaham }1979]{maz79}Mazeh T., 
Shaham J., 1979, A\&A, 77, 145 
\bibitem[\protect\citename{Paper I}]{pap1}Mazeh T., Krymolowski Y., 
Latham D. W., 1993, MNRAS, 263, 775 (Paper I)
\bibitem[\protect\citename{Mazeh, Krymolowski \& Rosenfeld
}1997]{maz97}Mazeh T., Krymolowski Y., Rosenfeld G., 1997, ApJ, 477,
L103
\bibitem[\protect\citename{Mazeh \& Mayor }1983]{maz83}Mazeh, T.,
Mayor, M., 1983, in Current Techniques in Double and Multiple Star
Research, I.A.U. Coll. No. 62, eds. R.S. Harrington and O.G. Franz,
Publ. of the Lowell Observatory; IX; 120
\bibitem[\protect\citename{Saar, N\"{o}rdstrom \& Andersen 
}1990]{saa90}Saar S. H., N\"{o}rdstrom B., Andersen J., 1990, A\&A, 
235, 291 
\bibitem[\protect\citename{Scargle }1982]{sca82}Scargle J. D., 1982, 
ApJ, 263, 835 
\bibitem[\protect\citename{S\"{o}derhjelm }1984]{sod84}S\"{o}derhjelm 
S., 1984, A\&A, 141, 232 
\bibitem[\protect\citename{Stefanik et al. }1997]{ste97}Stefanik R. 
P., Caruso J. R., Torres G., Jha S., Latham D. W., 1997, Baltic 
Astronomy, 6, 137 
\bibitem[\protect\citename{Tokovinin }1997]{tok97}Tokovinin A. A., 
1997, A\&AS, 124, 75 
\bibitem[\protect\citename{Tokovinin }1999]{tok99}Tokovinin A. A., 
1999, A\&AS, 136, 373 
\bibitem[\protect\citename{Upgren \& Rubin }1965]{upg65}Upgren A. R., 
Rubin V. C., 1965, PASP, 77, 355 
\bibitem[\protect\citename{Upgren et al. }1982]{upg82}Upgren A. 
R., Philip A. G. D., Beavers W. I., 1982, PASP, 94, 229 
\bibitem[\protect\citename{Zahn} 1975]{zah75}Zahn, J. P., 1975,
A\&A, 41, 329
\bibitem[\protect\citename{Zucker \& Mazeh }1994]{zuc94}Zucker S., 
Mazeh T., 1994, ApJ, 420, 806 
\bibitem[\protect\citename{Zucker, Torres \& Mazeh }1995]{zuc95}Zucker 
S., Torres G., Mazeh T., 1995, ApJ, 452, 863 
\end{thebibliography}
\end{document}